\documentclass[superscriptaddress,twocolumn,nofootinbib,aps,prx]{revtex4-2} 
\usepackage[utf8]{inputenc}
\usepackage{comment}
\usepackage{graphicx} 
\usepackage[section]{placeins}
\usepackage{mathtools} 
\usepackage{physics} 
\usepackage{amsmath,dsfont,physics} 
\usepackage{bm}
\usepackage{stmaryrd}
\usepackage{floatrow}
\usepackage[margin=1in]{geometry}
\usepackage[title]{appendix} 
\usepackage{blindtext} 
\usepackage{fancyhdr} 
\usepackage{indentfirst} 
\usepackage{natbib} 
\usepackage{xcolor}
\usepackage{newunicodechar}
\newunicodechar{ }{\,} 
\usepackage{txfonts}
\usepackage[export]{adjustbox}
\usepackage[colorlinks=true, linkcolor=blue, citecolor=blue, urlcolor=blue]{hyperref}
\usepackage[label font=bf,labelformat=simple,caption=false]{subfig}
\begin{document}

\title{Angular Momentum Fluctuations in the Phonon Vacuum of Symmetric Crystals}
\author{R.Yi}
\email{yira@bc.edu }
\affiliation{Department of Physics, Boston College, 140 Commonwealth Avenue, Chestnut Hill, Massachusetts 02467, USA}
\author{V. Williams}
\affiliation{Department of Physics, Boston College, 140 Commonwealth Avenue, Chestnut Hill, Massachusetts 02467, USA}

\author{B. Flebus}
\affiliation{Department of Physics, Boston College, 140 Commonwealth Avenue, Chestnut Hill, Massachusetts 02467, USA}

\date{\today}

\begin{abstract}
Although time-reversal and inversion symmetry constrain the angular momentum of each phonon mode to vanish, we show that the vacuum state of crystals with such symmetries can nevertheless exhibit finite angular momentum fluctuations, {which persist at finite temperature}. These fluctuations arise from quantum coherence between nondegenerate modes with \textcolor{black}{noncollinear polarizations} and are encoded in the off-diagonal components of the angular momentum operator. Their origin lies in the noncommutativity between the phonon Hamiltonian and angular momentum, which enables time-dependent rotational dynamics even in symmetric vacua. Using a minimal model, we provide an intuitive picture of this phenomenon in terms of beating between linearly polarized modes, which generates a finite instantaneous angular momentum while remaining symmetry-forbidden in the mean. \textcolor{black}{We further show that these vacuum fluctuations give rise to distinct finite-frequency spectral signatures and outline a concrete route for their detection using time-resolved spectroscopic probes sensitive to lattice polarization and symmetry. Our results identify a previously unexplored regime of lattice dynamics, revealing that even the symmetric phonon vacuum can harbor structured, dynamical angular-momentum correlations.}
\end{abstract}

\maketitle

\textit{Introduction.} 
\textcolor{black}{Vacuum fluctuations are among the most paradigmatic manifestations of quantum mechanics~\cite{milonni2013quantum}. Rooted in the noncommuting operator structure of quantum observables, they impose an irreducible noise floor that persists even at zero temperature.
 Their physical consequences are both profound and measurable,  underpinning  phenomena ranging from the irreducible zero-point motion of confined particles~\cite{heisenberg1985quantentheoretische}, to radiative shifts in atomic spectra~\cite{lamb1947fine,bethe1947electromagnetic,dirac1927quantum} and forces arising from vacuum stress in confined geometries~\cite{casimir1948attraction}.}

\textcolor{black}{These landmark examples have motivated a broader search for fluctuation-driven phenomena in other quantum settings, particularly as new degrees of freedom come into focus.  In crystalline solids, a natural candidate is phonon angular momentum—the orbital counterpart of spin in lattice vibrations—whose identification has catalyzed the rapidly developing field of  \textit{chiral phononics}~\cite{zhang2014angular,zhang2015chiral,Dynamicalmultiferroicity2017,Nova2017EffectiveMagneticFieldPhonons,OAMphonon2019,juraschek2025chiral}. Over the past decade, theory and experiment have established that phonons can carry angular momentum and generate Hall-like responses across a wide range of materials~\cite{Garanin2015,Kohno2018,Hamada2018,zhang2018Chiral,Park2020,zhangMeasurementPhononAngular2025,Cheng2020LargeEffectivePhononMagneticMoment,Wu2023FluctuationEnhancedPhononMagneticMoments,Zhang2023,phononHallViscosity2023,Sun2025HelixPAM,coh2023classification}. To date, however, these efforts have focused almost exclusively on systems that break time-reversal ($\mathcal{T}$) or inversion ($\mathcal{P}$) symmetry, where individual modes—or the lattice as a whole—can acquire a finite \emph{mean} angular momentum, and on finite-temperature responses driven by  thermal populations}.

\textcolor{black}{Here we address a complementary and previously unexplored regime—the phonon vacuum of a $\mathcal{PT}$-symmetric crystal—and show that it can exhibit finite angular-momentum fluctuations even though each individual phonon mode carries zero angular momentum. We find that  dynamical angular-momentum response is enabled by phonon pairs with noncollinear polarizations and nondegenerate frequencies—a condition generically realized in crystalline solids. We show explicitly that nondegeneracy lifts the continuous rotational freedom within the polarization subspace: as a result, the angular-momentum operator no longer commutes with the phonon Hamiltonian, and  the vacuum can exhibit  time-dependent rotational fluctuations.}

\begin{figure}[b!]
    \centering    
    \includegraphics[trim={4cm 0.5cm 4cm 2cm},clip,width=1\linewidth]{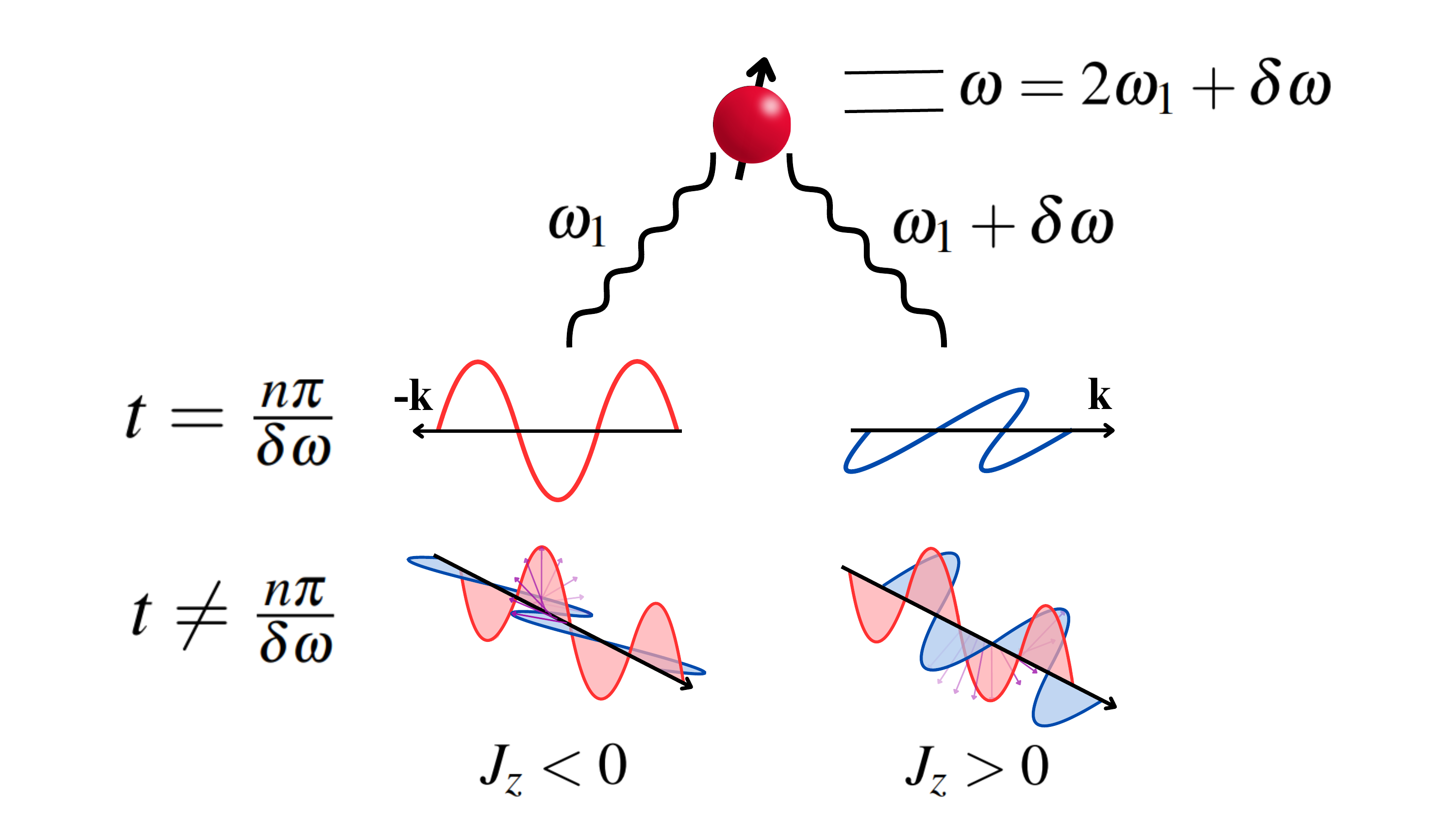}
  \caption{A two-level system with resonance frequency \( \omega \), coupled to vacuum angular momentum noise, undergoes spontaneous emission by generating a pair of phonons with orthogonal, linear polarizations which satisfy conservation of energy and linear momentum. The  angular momentum fluctuations associated with the phonon pair can be understood through a classical analogy: the superposition of two orthogonal, linearly polarized waves with a frequency mismatch \( \delta\omega = \omega_2 - \omega_1 \), which results into a finite instantaneous angular momentum $J_z$ that averages to zero over a period \( T= 2\pi / \delta\omega \).}
\label{fig:master}
\end{figure}


To provide physical intuition, we consider a minimal two-level system coupled to fluctuations of lattice angular momentum.
 As shown in Fig.~\ref{fig:master}, the coupling opens a spontaneous decay channel into the lattice: the two-level system relaxes by generating pairs of phonons with opposite momenta and orthogonal, linear  polarizations. From a classical perspective, the superposition of two orthogonal, linearly polarized waves with slightly different frequencies traces a slowly precessing elliptical trajectory: the instantaneous motion carries angular momentum, yet its time average vanishes. In our theory, the same dynamical “beating” structure is carried by the lattice zero-point motion: coherence between noncollinear, nondegenerate phonon modes—captured by off-diagonal matrix elements of the angular-momentum operator in the phonon eigenbasis—produces angular-momentum fluctuations with zero mean but finite variance.
\textcolor{black}{These vacuum fluctuations are directly measurable: we outline an experimentally feasible pump–probe scheme, based on resonant two-phonon Raman excitation and time-resolved optical polarimetry, taking silicon as a benchmark.}

Our work opens a new direction for probing vacuum structure in crystalline solids, uncovering a class of symmetry-protected quantum correlations in phonon systems and laying the foundation for exploring their impact on angular momentum transport, quantum geometry, and lattice-mediated spin coherence.

\textcolor{black}{\textit{Angular momentum fluctuations.}}
We start from a generic phonon Hamiltonian of the form
\begin{equation}\label{eq1}
\hat{H} = \hbar \sum_{\mathbf{k}, \sigma} \omega_{\mathbf{k}, \sigma} \left( \hat a_{\mathbf{k}, \sigma}^\dagger \hat a_{\mathbf{k}, \sigma} + \frac{1}{2} \right)\,,
\end{equation}
where \( a_{\mathbf{k}, \sigma}^\dagger \) denotes the creation operator of a phonon mode with wavevector \( \mathbf{k} \) and branch index \( \sigma \). For concreteness, here we focus on the $z$-component of the phonon angular momentum, i.e.,
\begin{equation}\label{eq:Jz}
\hat{J}_z^{\,\rm ph} = \sum_l \hat{\mathbf{u}}_l^T (i M)\, \dot{\hat{\mathbf{u}}}_l, \quad \text{with} \quad M\equiv \begin{pmatrix} 0 & -i&0 \\ i & 0&0\\0&0&0 \end{pmatrix},
\end{equation}
while extension to other components is straightforward~\cite{zhang2014angular,zhang2015chiral}.  Employing the standard normal-mode expansion of the $l$th unit-cell displacement operator $\hat{\mathbf u}_l$, Eq.~(\ref{eq:Jz}) can be rewritten as
 \begin{equation}\label{eq3}
\hat J_z^{\,\rm ph}
=\hat J_{\rm nc}
+\frac{\hbar}{2}\sum_{\mathbf{k},\sigma,\sigma'}\left (F_{\mathbf{k},\sigma,\sigma'}
\,\hat a_{\mathbf{k},\sigma}^\dagger \hat a_{-\mathbf{k},\sigma'}^\dagger
+\mathrm{h.c.}\right)\,,
\end{equation}

where  the number-conserving term reads as
\[
\begin{aligned} \hat J_\text{nc}= & \frac{\hbar}{2} \sum_{\mathbf{k}, \sigma, \sigma^{\prime}} \sqrt{\frac{\omega_{\mathbf{k}, \sigma^{\prime}}}{\omega_{\mathbf{k}, \sigma}}}\left[\varepsilon_{\mathbf{k}, \sigma}^{\dagger} M \varepsilon_{\mathbf{k}, \sigma^{\prime}} \hat a_{\mathbf{k}, \sigma}^{\dagger} \hat a_{\mathbf{k}, \sigma^{\prime}}\right. \\&\left.-\varepsilon_{\mathbf{k}, \sigma}^T M \varepsilon_{\mathbf{k}, \sigma^{\prime}}^*\left(\delta_{\sigma ,\sigma^{\prime}}+\hat a_{\mathbf{k}, \sigma^{\prime}}^{\dagger} \hat a_{\mathbf{k}, \sigma}\right)\right] \,, \end{aligned}
\]
while the coefficient of the anomalous terms is
\[
\begin{aligned}
F_{\mathbf{k},\sigma,\sigma'}
\equiv\tfrac12\Bigl(\sqrt{\tfrac{\omega_{\mathbf{k},\sigma}}{\omega_{\mathbf{k},\sigma'}}}
-\sqrt{\tfrac{\omega_{\mathbf{k},\sigma'}}{\omega_{\mathbf{k},\sigma}}}\Bigr)
\,\varepsilon_{\mathbf{k},\sigma}^\dagger M\,\varepsilon_{-\mathbf{k},\sigma'}^*\,,
\end{aligned}
\]
where $\varepsilon_{\mathbf{k},\sigma}$ is a displacement polarization vector.

The expectation value of \(\hat J_z^{\,\text{ph}}\)~\eqref{eq3} takes the conventional form:
\begin{align}
\langle \hat J_z^{\,\mathrm{ph}}\rangle=\sum_{\mathbf{k}, \sigma} \ell^z_{\mathbf{k}, \sigma}\left[f\left(\omega_{\mathbf{k}, \sigma}\right)+\frac{1}{2}\right], \; \; \ell^z_{\mathbf{k}, \sigma}\equiv \hbar\,\varepsilon_{\mathbf{k},\sigma}^\dagger M\,\varepsilon_{\mathbf{k},\sigma}\,,
\label{eq4}
\end{align}

where \( f( \omega_{\mathbf{k}, \sigma}) = \left(e^{ \hbar \omega_{\mathbf{k}, \sigma} / k_B T} - 1\right)^{-1} \) is the Bose--Einstein distribution function, and \( \ell^z_{\mathbf{k}, \sigma} \) is the angular momentum of the $(\textbf{k}, \sigma$) phonon mode.
In $\mathcal{PT}$-symmetric crystals, each polarization vector \( \varepsilon_{\mathbf{k}, \sigma} \) can be chosen to be real, which, in turn, constrains both the individual and total phonon angular momentum to vanish, i.e., \( \ell^z_{\mathbf{k}, \sigma}=0, \langle \hat{J}_z^{\,\mathrm{ph}} \rangle = 0 \)~\cite{coh2023classification}. The number-nonconserving terms proportional to $F_{\mathbf{k},\sigma,\sigma'}$ in Eq.~(\ref{eq3}) oscillate at $\omega_{\mathbf{k},\sigma}+\omega_{\mathbf{k},\sigma'}$ and therefore do not contribute to  the  angular momentum mean; \textcolor{black}{however, as we will show below, they are precisely the source of  vacuum angular-momentum fluctuations with  finite variance.}

To quantify angular-momentum noise, we introduce the retarded susceptibility of the vacuum state $| 0 \rangle$ as
\begin{equation}\label{eq5}
\chi(\omega)
= -\frac{i}{\hbar} \int_0^\infty dt\, e^{i(\omega+i0^+)t}
\langle 0 |[\hat{J}_z^{\, \mathrm{ph}}(t), \hat{J}_z^{\,\mathrm{ph}}(0)] | 0 \rangle\,,
\end{equation}
where $0^+$ enforces causality and convergence~\cite{SM}.   \textcolor{black}{Evaluating Eq.~\eqref{eq5} using Wick’s theorem~\cite{Wick}, we find that the only nonvanishing contributions to the dynamical response arise from pair-creation processes followed by pair annihilation at later times, generated by the anomalous (number-nonconserving) terms in Eq.~\eqref{eq3}}~\cite{SM}.   After isolating the nonvanishing operator combinations, we obtain

\begin{align}
\langle[\hat J_z^{\,\text{ph}}(t),\hat J_z^{\,\text{ph}}(0)]\rangle=&\frac{\hbar ^2}{4}\sum_{\textbf{k},\sigma,\sigma'}\Bigl[ e^{-i(\omega_{\textbf{k},\sigma}+\omega_{-\textbf{k},\sigma'})t}F^*_{\textbf{k},\sigma,\sigma'}\notag\\
&\times(F_{\textbf{k},\sigma,\sigma'}+F_{-\textbf{k},\sigma',\sigma}) -\mathrm{c.c.}\Bigr]\,,
\label{eq6}
\end{align}

In the presence of \( \mathcal{PT} \) symmetry, Eq.~(\ref{eq6}) further simplifies under the constraints \( \omega_{\mathbf{k},\sigma} = \omega_{-\mathbf{k},\sigma} \) and \( F_{\mathbf{k},\sigma,\sigma'} = F_{-\mathbf{k},\sigma',\sigma} \).
Equation~\eqref{eq5} can then be written as

\begin{align}
\chi(\omega)
=&\frac{\hbar }{4}\sum_{\mathbf{k},\sigma\neq\sigma'}
|\varepsilon_{\mathbf{k},\sigma}^T M\,\varepsilon_{-\mathbf{k},\sigma'}|^2\frac{(\omega_{\mathbf{k},\sigma}-\omega_{\mathbf{k},\sigma'})^2}
     {\omega_{\mathbf{k},\sigma}\omega_{\mathbf{k},\sigma'}}\notag\\
&\times
\frac{\omega_{\mathbf{k},\sigma}+\omega_{\mathbf{k},\sigma'}}{(\omega+i0^+)^2-(\omega_{\mathbf{k},\sigma}+\omega_{\mathbf{k},\sigma'})^2}\,.
\label{eq7}
\end{align}

Equation~\eqref{eq7}, which represents the central result of this work, shows that vacuum fluctuations of angular momentum exhibit a finite spectral weight due to off-diagonal coherences  between phonons with noncollinear polarizations and nondegenerate frequencies.  \textcolor{black}{The mode-resolved weight is nonnegative and vanishes as the two branches approach degeneracy, i.e., for 
$\delta\omega= |\omega_{\mathbf{k},\sigma}-\omega_{\mathbf{k},\sigma'}| \ll \omega_{\mathbf{k},\sigma}$, 
one finds $(\omega_{\mathbf{k},\sigma}-\omega_{\mathbf{k},\sigma'})^2/
(\omega_{\mathbf{k},\sigma}\omega_{\mathbf{k},\sigma'})
\sim (\delta\omega/\omega_{\mathbf{k},\sigma})^2$. The imaginary component of the dynamical susceptibility $\chi''(\omega)\equiv-{\rm Im}\,\chi(\omega)$ satisfies $\chi''(\omega>0)\ge0$, as required by causality, and admits a direct physical interpretation as a two-phonon absorption spectrum driven by zero-point angular-momentum fluctuations~\cite{SM}.} 
\textcolor{black}{
The dynamical response is thus controlled by $\mathbf{k}$-points that simultaneously maximize (i) the two-phonon density of states and (ii) the off-diagonal angular-momentum coherence 
$\big|\varepsilon_{\mathbf{k},\sigma}^{T} M\,\varepsilon_{-\mathbf{k},\sigma'}\big|^{2}$ 
between two nondegenerate modes.
In a minimal 2$d$ model, the coherence  saturates its upper bound for $\sigma\neq\sigma'$, i.e., $\big|\varepsilon_{\mathbf{k},\sigma}^{T} M\,\varepsilon_{-\mathbf{k},\sigma'}\big|^{2}=1-\delta_{\sigma\sigma'}$, whereas, in a  3$d$ crystal, it remains of order one only when the two modes are related by an in-plane rotation, i.e., when they form orthogonal polarization components of the same atomic motion. As a concrete example, at the $\mathrm{X}$ point of $\mathcal{PT}$-symmetric silicon the LO ($\parallel\hat{\mathbf{x}}$) and TO ($\parallel\hat{\mathbf{y}}$) modes are directly mixed by $M$ [$M\hat{\mathbf{x}}\propto\hat{\mathbf{y}}$, $M\hat{\mathbf{y}}\propto-\hat{\mathbf{x}}$], yielding $\big|\varepsilon_{\mathbf{k},\mathrm{LO}}^{T} M\,\varepsilon_{-\mathbf{k},\mathrm{TO}}\big|^{2}\sim\mathcal{O}(1)$~\cite{SM}. By contrast,  the off-diagonal coherence between modes polarized along the same direction,  e.g.,  LA ($\parallel\hat{\mathbf{x}}$) and LO ($\parallel\hat{\mathbf{x}}$), is parametrically suppressed.} 

\begin{figure*}[htbp]
    \centering
\includegraphics[width=1\linewidth]{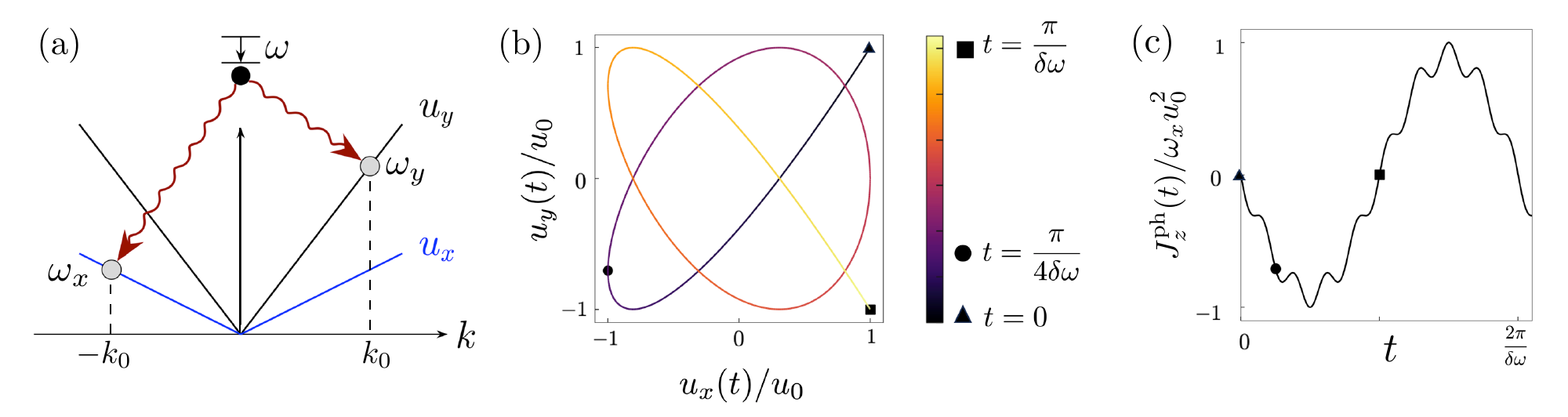}
    \caption{(a) Schematic illustration of a two-level system with resonance frequency $\omega$ undergoing spontaneous emission into a two-dimensional, $\mathcal{PT}$-symmetric crystal. As dictated by Eq.~\eqref{eq:susceptibility}, the emission produces a pair of phonons with opposite linear momenta $\pm k_0$ and orthogonal polarizations, taken here along $\hat{\mathbf{x}}$ and $\hat{\mathbf{y}}$. Because the two branches are nondegenerate, their frequencies differ : the branch with the larger group velocity has the higher frequency, $\omega_y>\omega_x$, while energy conservation enforces $\omega=\omega_x+\omega_y$. The resulting two-phonon state admits a classical representation as a coherent superposition of orthogonal displacement fields $u_x(t)$ and $u_y(t)$, oscillating at frequencies $\omega_x$ and $\omega_y$, respectively.
(b) Classical trajectory traced by the displacement field~\eqref{superpo}, plotted as $u_y(t)/u_0$ versus $u_x(t)/u_0$ over half a beating cycle ($T/2=\pi/\delta\omega$), for $\omega_x=1$ and detuning $\delta\omega=\omega_y-\omega_x=0.25$. The resulting elliptical motion is color-coded by time, with representative time instants indicated to highlight the precession of the polarization ellipse.
 (c) Time evolution of the instantaneous phonon angular momentum in Eq.~\eqref{eq9}, shown in dimensionless form as $J_z^{\rm ph}(t)/(\omega_x u_0^2)$ over one full period $T$. The oscillations originate from the beating dynamics in (b) and have finite amplitude but zero time average, mirroring the structured temporal fluctuations captured by Eq.~\eqref{comm}.}
    \label{fig:1}
\end{figure*}

\textcolor{black}{\textit{Fluctuation dynamics.}} The generation of angular momentum fluctuations can be understood intuitively by considering a quantum two-level spin \( \boldsymbol{\sigma} \) with resonance frequency \( \omega \), coupled to the angular momentum of the lattice via an interaction of the form \( \lambda \sigma^{+} \hat{J}^{\text {ph}}_z + \text{h.c.} \), with \( \sigma^{\pm} = (\sigma^x \pm i \sigma^y)/2 \) and $\lambda$ set by microscopic details. The spontaneous emission rate of the quantum spin follows from Fermi’s golden rule~\cite{SM}: 

\begin{equation}
\Gamma(\omega)=\pi\,|\lambda|^2\!\sum_{\mathbf{k},\sigma\neq\sigma'}
\Bigl|F_{\mathbf{k},\sigma,\sigma'}\Bigr|^2
\delta\bigl(\omega-\omega_{\mathbf{k},\sigma}-\omega_{\mathbf{k},\sigma'}\bigr).
\label{eq:susceptibility}
\end{equation}
\textcolor{black}{which describes spin relaxation via the resonant emission of a phonon pair with opposite momenta \( \mathbf{k} \) and \( -\mathbf{k} \), whose likelihood is enhanced when the two modes are  orthogonally polarized and significantly nondegenerate. These same conditions imply that the emitted pair is created in a polarization-entangled, phase-coherent superposition at fixed wavevector, rather than as an incoherent mixture of independent phonons.
As a consequence, the coherent superposition of orthogonally polarized, nondegenerate modes acquires a nontrivial internal time evolution, which manifests as a nonzero instantaneous angular momentum. }
 This intuition can be confirmed formally by evaluating the commutator~\cite{SM}:

\begin{align}
&\left[\hat J_z^{\,\rm ph}, \hat{H}\right]=\frac{\hbar^2}{2} \sum_{\mathbf k, \sigma\neq\sigma^{\prime}} \varepsilon_{\mathbf{k}, \sigma}^{T} M \varepsilon_{\mathbf{k} ,\sigma^{\prime}}\frac{\omega^2_{\mathbf{k},\sigma'}-\omega^2_{\mathbf{k},\sigma}}{\sqrt{{\omega_{\mathbf{k},\sigma}\omega_{\mathbf{k},\sigma'}}}}\notag\\
&\times \left(\hat a_{\mathbf{k} ,\sigma}^{\dagger} \hat a_{\mathbf{k}, \sigma^{\prime}}+\frac{1}{2}\hat a_{\mathbf{k},\sigma}^{\dagger} \hat a^\dagger_{\mathbf{-k} ,\sigma^{\prime}}-\frac{1}{2}\hat a_{\mathbf{k},\sigma}\hat a_{\mathbf{-k} ,\sigma^{\prime}}\right)\, ,
\label{comm}
\end{align}

which shows that nondegeneracy between phonon branches introduces a spectral asymmetry that explicitly breaks in-plane polarization-rotation symmetry,  endowing the vacuum with an instantaneous, zero-mean handedness {and supplying the fluctuating torque that feeds the two-phonon continuum.}
In contrast, when the branches are degenerate, the phonon modes can be combined into circularly polarized states that are simultaneously eigenstates of both the Hamiltonian and the angular momentum operator. These states carry finite angular momentum, but remain stationary and thus do not exhibit temporal fluctuations.

While our results are inherently quantum mechanical, drawing an analogy with a coherent superposition of two classical, orthogonal linearly polarized waves, \( u_x(t) \) and \( u_y(t) \), provides clear physical insight into how a frequency splitting between modes leads to angular momentum fluctuations.
As illustrated in Fig.~\ref{fig:1}(a), the conservation of energy and momentum dictated by Eq.~\eqref{eq:susceptibility} selects a particular polarization structure for the emitted phonon pair. Specifically, for a given \( \mathbf{k} \) and \( \omega \), energetically allowed coherent superpositions favor configurations in which the phonon mode with higher group velocity carries a larger share of the total energy. Assuming that the \( u_x \) field propagates more slowly than \( u_y \), the analogous classical displacement field takes the form
\begin{equation}
\mathbf{u}(t) = \mathrm{Re}\left[u_0e^{-i\omega_x t} \left( \hat{\mathbf{x}} + \hat{\mathbf{y}} \, e^{-i\delta\omega t} \right)\right],
\label{superpo}
\end{equation}
where \( \omega_x \) and \( \omega_y = \omega_x + \delta\omega \), with \( \delta\omega > 0 \), denote the frequencies of the classical waves \( u_x(t) \) and \( u_y(t) \), respectively, and \( u_0 \) is a real, positive constant setting the amplitude of the displacement field.
 Figure~\ref{fig:1}(b) shows that the coherent superposition~\eqref{superpo} traces a time-evolving elliptical trajectory in the \( xy \)-plane, with the relative phase \( \delta\omega \cdot t \) modulating the instantaneous ellipticity of the displacement field.  
  The instantaneous angular momentum \( J_z^{\,\rm ph}(t) = u_x \dot{u}_y - u_y \dot{u}_x \) generated during the time evolution can be straightforwardly calculated as
\begin{equation}\label{eq9}
\frac{J_z^{\,\rm ph}(t)}{u_0^2} = -\frac{\delta\omega}{2} \sin\left[\left(2 \omega_x + \delta\omega\right)t\right] - \frac{2 \omega_x + \delta\omega}{2} \sin(\delta\omega t),
\end{equation}

{which captures envelope–carrier beating between two nondegenerate polarizations. As shown by Fig.~\ref{fig:1}(c),  $J_z^{\,\rm ph}(t)$ displays finite-amplitude, zero-mean oscillations with period $T=2\pi/\delta\omega$ and fast carrier oscillations at frequency $2\omega_x+\delta\omega$. Equation~\eqref{eq9} can be interpreted as the coherent-state projection of the operator precession generated by Eq.~\eqref{comm}:
 the number-conserving coherence $\hat a_{\mathbf{k},\sigma}^{\dagger}\hat a_{\mathbf{k},\sigma'}$ evolves at  frequency $\lvert \omega_{\mathbf{k},\sigma}-\omega_{\mathbf{k},\sigma'} \rvert$, whereas the anomalous pair $\hat a_{\mathbf{k},\sigma}^{\dagger}\hat a_{-\mathbf{k},\sigma'}^{\dagger}$ precesses at frequency $\omega_{\mathbf{k},\sigma}+\omega_{\mathbf{k},\sigma'}$. When evaluated on the coherent two-mode state described by Eq.~\eqref{superpo}, these contributions yield the slow envelope at $\delta \omega$ and the fast carrier at  \( 2\omega_x + \delta\omega \) entering Eq.~\eqref{eq9}}~\cite{SM}. 

 This two-tone structure is absent in Eq.~\eqref{eq:susceptibility} because only the number-nonconserving coherence contributes to the  vacuum commutator. At finite temperature, however, thermal populations activate the normal, number-conserving channel, adding spectral weight at \(|\omega_{\mathbf{k},\sigma}-\omega_{\mathbf{k}',\sigma'}|\) and thereby restoring the two-frequency response~\cite{SM}.
 





\textit{Conclusion and outlook.} \textcolor{black}{ In this work, we show that the phonon vacuum of a $\mathcal{PT}$-symmetric crystal hosts finite angular-momentum fluctuations, even though the mean phonon angular momentum is symmetry-forbidden. In contrast to mechanisms that invoke intrinsic or extrinsic symmetry breaking to produce chiral phonons with nonzero angular momentum~\cite{flebus2022charged,coh2023classification,xue2025extrinsic}, the effect we uncover is purely coherence-driven and controlled by the off-diagonal matrix elements of the phonon angular-momentum operator.} 
The key ingredient is the presence of two nondegenerate phonon modes whose polarization are not collinear  --  a condition realized in a broad class of crystalline materials that leads to the noncommutativity between the phonon Hamiltonian and angular momentum operator. 

\textcolor{black}{We compute the dynamical response of the phonon vacuum and find that its spectral weight tracks the two-phonon density of states, with dominant contributions from nondegenerate, orthogonally polarized branches. This response is purely AC and independent of thermal occupation, therefore fundamentally distinct from DC phonon thermal Hall effects. Spontaneous emission  provides a natural probe: finite-frequency vacuum fluctuations can, in principle, allow a quantum system coupled to phonon angular momentum to relax by emitting correlated phonon pairs.  The emitted state carries inter-branch coherence between the two branches, converting the motion into an oscillatory chiral lattice trajectory with finite instantaneous angular momentum—in close analogy to the polarization beating of two nondegenerate, orthogonally polarized classical waves.} 

\textcolor{black}{The coherence underlying the vacuum response can be also interrogated without relying on spontaneous emission, i.e., coherently preparing an appropriate two-mode superposition and monitoring the resulting oscillatory axial signal can provide a direct, time-resolved probe of a response that is symmetry-forbidden in the mean.}
In this context, a natural probe is time-resolved resonant two-phonon Raman spectroscopy~\cite{klein1983resonant,ruhman1987timeresolved} . At electronic resonance, an ultrafast pump can prepare correlated, nondegenerate phonon pairs with opposite wavevectors via the second‑order Raman pathway. \textcolor{black}{The Raman-driven pair coherence couples directly to the axial channel encoded in the off-diagonal angular-momentum matrix elements. Although this channel is symmetry-forbidden in equilibrium in a $\mathcal{PT}$-symmetric crystal, it can appear transiently as an antisymmetric modulation of the dielectric response, observable as time-dependent probe polarization rotation or ellipticity in a Faraday/Kerr-type geometry ~\cite{basini2024dynamical,davies2024phononic,chen2015helicity,fu2020anomalous,luo2023chiral,parlak2023helicity,OAMphonon2019, ji2016giant, fiorazzo2025theory,merlin2024unraveling}. Such signal can be isolated by suppressing the dominant symmetric background with crossed linear pump–probe polarizations and then projecting onto the axial component with a right–left circular differential readout.}

The dynamical angular-momentum fluctuations identified here arise when two phonon branches with opposite linear momentum are nondegenerate in energy and noncollinear in polarization --- a situation generically realized for both acoustic and optical phonons.
 While acoustic modes  offer a natural window into these fluctuations at finite temperature, the vacuum limit is most cleanly approached with optical phonons: their higher frequencies strongly suppress thermal populations and allow near-ground-state initialization without cryogenic cooling.
As a concrete platform, we propose silicon, which hosts nondegenerate zone-edge LO and TO optical modes with orthogonal polarizations at the $X$ points, together with a strong resonant enhancement of the two-phonon Raman density of states~\cite{klein1974selective}.
 The associated inter-branch coherence would drive an angular-momentum oscillation at $\delta\omega \simeq 1.6~\mathrm{THz}$~\cite{SM,Johnson1959}, i.e., resulting into a sub-picosecond beat period easily accessible to ultrafast pump--probe polarimetry~\cite{kirilyuk2010ultrafast}.

Beyond their experimental accessibility, our findings uncover a previously unrecognized dimension of lattice dynamics, with the potential to shed new light on the dynamical and geometric properties of even the simplest crystalline materials.

\textit{Acknowledgments.}---The authors thank Q. Niu, D. Bossini and J. M. P. Nair for insightful discussions, and E. Foisy for the production of illustrations. B. Flebus acknowledges support from DOE under Grant. No DE- SC0024090.

\bibliography{references}


\end{document}


\title{Supplemental Material for:\\[2pt]
\emph{Angular Momentum Fluctuations in the Phonon Vacuum of Symmetric Crystals}}

\author{R.\ Yi}
\email{yira@bc.edu}
\affiliation{Department of Physics, Boston College, 140 Commonwealth Avenue, Chestnut Hill, Massachusetts 02467, USA}

\author{V.\ Williams}
\affiliation{Department of Physics, Boston College, 140 Commonwealth Avenue, Chestnut Hill, Massachusetts 02467, USA}

\author{B.\ Flebus}
\affiliation{Department of Physics, Boston College, 140 Commonwealth Avenue, Chestnut Hill, Massachusetts 02467, USA}

\date{\today}

\maketitle

This Supplemental Material is organized as follows. 
Section~S1 starts from the linear-response definition [Eqs.~(\textcolor{blue}{5})–(\textcolor{blue}{6}) of the main text] 
and derives the closed-form expression in Eq.~(\textcolor{blue}{7}), including its finite-temperature generalization. 
Section~S2 applies the formalism to a two-level spin coupled to the phonon angular momentum 
$J_z^{\mathrm{ph}}$, derives the relaxation rate quoted in Eq.~(\textcolor{blue}{8}) (with a 2D illustrative model), 
and extends the result to finite temperature. 
Section~S3 connects the commutator representation in Eq.~(\textcolor{blue}{9}) to a coherent two-mode construction 
that reproduces Eqs.~(\textcolor{blue}{10})–(\textcolor{blue}{11}) and clarifies the carrier–envelope structure. Section S4 demonstrates in silicon that resonant two-phonon Raman excitation of nondegenerate LO and TO modes at $±X$ can generate instantaneous lattice angular momentum, potentially detectable as time-dependent probe polarization rotation/ellipticity signals.

\section{S1.  Phonon angular momentum susceptibility}
\setcounter{equation}{0}

We define the (causal) susceptibility associated with fluctuations of the $z$-component of the phonon angular momentum operator as
\begin{equation}
\chi(\omega)
= -\frac{i}{\hbar}\int_{0}^{\infty}\!dt\,e^{i(\omega+i0^+)t}\,
\big\langle[\hat J^{\mathrm{ph}}_{z}(t), \hat J^{\mathrm{ph}}_{z}(0) ]\big\rangle \,,
\label{eq:Kubo}
\end{equation}
where $\langle\cdots\rangle$ denotes the expectation value in a given stationary state, while the infinitesimal $+i0^+$ enforces causality and convergence.
It is convenient to recast Eq.~\eqref{eq:Kubo} as 
\begin{equation}
\chi (\omega)
=\chi _{\rm a}(\omega)+\chi _{\rm nc}(\omega)\,,
\end{equation}
where

\begin{align}
\chi_{\mathrm{a}}(\omega)
&= -\frac{i}{\hbar}\int_{0}^{\infty}\!dt\,e^{i(\omega+i0^+) t}\,
\langle [\,\hat{J}_{\mathrm{a}}(t),\,\hat{J}_{\mathrm{a}}(0)\,]\rangle\,,
\label{Chia}
\end{align}
and
\begin{align}
\chi_{\mathrm{nc}}(\omega)
&= -\frac{i}{\hbar}\int_{0}^{\infty}\!dt\,e^{i(\omega+i0^+) t}\,
\langle [\,\hat{J}_{\mathrm{nc}}(t),\,\hat{J}_{\mathrm{nc}}(0)\,]\rangle\,,
\label{Chinc}
\end{align}

correspond, respectively, to the susceptibility due to fluctuations of the anomalous, $\chi_{\mathrm{a}}$, and number-conserving, $\chi_{\mathrm{nc}}$, sectors of the phonon angular momentum operator defined in Eq.~(\textcolor{blue}{3}) of the main text. 

Operator products of the form $ \hat J_{\mathrm{nc}}\hat J_{\mathrm{a} }$ do not conserve phonon number and therefore do not contribute to Eq.~\eqref{eq:Kubo}.

\paragraph{Zero-Temperature Susceptibility.}
Evaluating four-operator averages by Wick’s theorem in the phonon vacuum, any normal-ordered string has vanishing expectation value. In particular,
\begin{align} &\langle 0|\,\hat a^{\dagger}_{\mathbf k,\sigma}\,\hat a_{\mathbf k,\sigma'}\,|0\rangle = 0 \,,\label{wick2}\\[-2pt] &\langle 0|\,\hat a^{\dagger}_{\mathbf k,\sigma}\,\hat a_{\mathbf k,\sigma'}\,\hat a^{\dagger}_{\mathbf k',\lambda}\,\hat a_{\mathbf k',\lambda'}\,|0\rangle = 0 \,,\label{wick4} \end{align}
since the only nonzero two-point contraction in the vacuum is
$
\langle 0|\,\hat a_{\mathbf k,\sigma}\,\hat a^{\dagger}_{\mathbf k',\lambda}\,|0\rangle
=\delta_{\mathbf k , \mathbf k'}\,\delta_{\sigma, \lambda}$ , 
and every contraction of the four-operator product leaves at least one annihilation operator acting on \(|0\rangle\).
From the definition of $\hat J_{\rm nc}$ in Eq.~(\textcolor{blue}{3}), its vacuum correlator is time-independent, i.e., 
 $\langle 0| \hat{J}_{\mathrm{nc}}(t) \hat{J}_{\mathrm{nc}}(0)|0\rangle=\langle 0| \hat{J}^2_{\mathrm{nc}}|0\rangle$, so the retarded commutator~\eqref{Chinc} — and hence the connected response — vanishes, i.e., $\chi _{\mathrm{nc}}(\omega)=0$. On the other hand, anomalous (pairing) processes, where the same two-phonon pair is created and later annihilated, yield a finite contribution to the susceptibility:
\begin{equation}
\begin{aligned}
\langle 0|\,\hat a_{\kk,\sigma}\hat a_{-\kk,\sigma'}\hat a_{\kk’,\lambda}^\dagger \hat a_{-\kk',\lambda'}^\dagger\,|0\rangle & = \delta_{\kk,\kk'}\delta_{\sigma,\lambda}\,\delta_{-\kk,-\kk'}\delta_{\sigma',\lambda'} \\
&+ \delta_{\kk,-\kk'}\delta_{\sigma,\lambda'}\,\delta_{-\kk,\kk'}\delta_{\sigma',\lambda}. 
\label{eq:Wick}
\end{aligned}
\end{equation}
Invoking Eq.~\eqref{eq:Wick}, reinstating the explicit time dependence $\hat a_{\mathbf{k},\sigma}(t)\hat a_{-\mathbf{k},\sigma'}(t)\!\propto\! e^{-i\left(\omega_{\mathbf{k}, \sigma}+\omega_{-\mathbf{k}, \sigma^{\prime}}\right) t}$, and performing the $\delta$-function sums,  the commutator in Eq.~\eqref{Chia} reduces to

\begin{equation}
\begin{aligned}
&\langle 0| \hat{J}_{\mathrm{a}}(t) \hat{J}_{\mathrm{a}}(0)|0\rangle= \frac{\hbar^2}{4} \sum_{\substack{\mathbf{k}, \sigma \neq \sigma^{\prime}  \\ \mathbf{k}^{\prime}, \lambda\neq\lambda^{\prime}}} F_{\mathbf{k}, \sigma, \sigma^{\prime}}^* F_{\mathbf{k}^{\prime}, \lambda, \lambda^{\prime}} \\
& \quad\times e^{-i\left(\omega_{\mathbf{k}, \sigma}+\omega_{-\mathbf{k}, \sigma^{\prime}}\right) t}\langle 0| \hat{a}_{-\mathbf{k}, \sigma^{\prime}} \hat{a}_{\mathbf{k}, \sigma} \hat{a}_{\mathbf{k}^{\prime}, \lambda}^{\dagger} \hat{a}_{-\mathbf{k}^{\prime}, \lambda^{\prime}}^{\dagger}|0\rangle \\
&= \frac{\hbar^2}{4} \sum_{\mathbf{k}, \sigma \neq \sigma^{\prime}} e^{-i\left(\omega_{\mathbf{k}, \sigma}+\omega_{-\mathbf{k}, \sigma^{\prime}}\right) t} F_{\mathbf{k}, \sigma, \sigma^{\prime}}^*\left[F_{\mathbf{k}, \sigma, \sigma^{\prime}}+F_{-\mathbf{k}, \sigma^{\prime}, \sigma}\right] \,, \\
& =\frac{\hbar^2}{4} \sum_{\mathbf{k}, \sigma \neq \sigma^{\prime}} \left[e^{-i\left(\omega_{\mathbf{k}, \sigma}+\omega_{-\mathbf{k}, \sigma^{\prime}}\right) t}-e^{i\left(\omega_{\mathbf{k}, \sigma}+\omega_{-\mathbf{k}, \sigma^{\prime}}\right) t} \right]\\
&\quad\times \,F_{\mathbf{k}, \sigma, \sigma^{\prime}}^*\left[F_{\mathbf{k}, \sigma, \sigma^{\prime}}+F_{-\mathbf{k}, \sigma^{\prime}, \sigma}\right]. 
\end{aligned}
\label{Jcommutator}
\end{equation}


Under combined \(\mathcal{P}\) and \(\mathcal{T}\) symmetry (\(\omega_{\mathbf{k},\sigma}=\omega_{-\mathbf{k},\sigma}\), \({\varepsilon}_{-\mathbf{k},\sigma}={\varepsilon}_{\mathbf{k},\sigma}^{*}\)), one finds \(F_{-\mathbf{k},\sigma',\sigma}=F_{\mathbf{k},\sigma,\sigma'}\). Consequently, Eq.~\eqref{Jcommutator} can be further simplified to
\begin{equation}
\begin{aligned}
\langle 0|[\Jph(t),\Jph(0)]|0\rangle
& =\frac{\hbar^2}{2} \sum_{\mathbf{k}, \sigma \neq \sigma^{\prime}}|F_{\kk,\sigma,\sigma'}|^2 \\
&\quad\times \,[e^{-i\left(\omega_{\mathbf{k}, \sigma}+\omega_{\mathbf{k}, \sigma^{\prime}}\right) t}-\mathrm{c.c.} ]. 
\label{eq:JJ-commutator}
\end{aligned}
\end{equation}
Substituting Eq.~\eqref{eq:JJ-commutator} into Eq.~\eqref{eq:Kubo} and performing the time integral gives the compact zero-temperature representation: 
\begin{equation}
\begin{aligned}
\ChiR(\omega)
= &\frac{\hbar}{2}\sum_{\kk,\sigma\neq\sigma'} |F_{\kk,\sigma,\sigma'}|^2
\left[\frac{1}{\omega-(\omega_{\kk,\sigma}+\omega_{\kk,\sigma'})+\ii 0^+}\right.\\
&\left.
-\,\frac{1}{\omega+(\omega_{\kk,\sigma}+\omega_{\kk,\sigma'})+\ii 0^+}\right]\,,
\label{eq:chiR-prekernel}
\end{aligned}
\end{equation}
where $\ChiR(\omega)=\chi_{\rm a}(\omega)$. Using
\begin{equation}
|F_{\kk,\sigma,\sigma'}|^2=
\frac{1}{4}\left(\frac{\omega_{\kk,\sigma}}{\omega_{\kk,\sigma'}}+\frac{\omega_{\kk,\sigma'}}{\omega_{\kk,\sigma}}-2\right)
\left|\eps_{\kk,\sigma}^{\TT} M\,\eps_{-\kk,\sigma'}\right|^2 \,,
\label{eq:kernel-def}
\end{equation}
one finds
\begin{equation}
\begin{aligned}
\chi(\omega)
&=\frac{\hbar }{4}\sum_{\mathbf{k},\sigma\neq\sigma'}
|\varepsilon_{\mathbf{k},\sigma}^T M\,\varepsilon_{-\mathbf{k},\sigma'}|^2\Bigl(\frac{\omega_{\mathbf{k},\sigma}}{\omega_{\mathbf{k},\sigma'}}+\frac{\omega_{\mathbf{k},\sigma'}}{\omega_{\mathbf{k},\sigma}}-2\Bigr)\\
&\times
\frac{\omega_{\mathbf{k},\sigma}+\omega_{\mathbf{k},\sigma'}}{(\omega+i0^+)^2-(\omega_{\mathbf{k},\sigma}+\omega_{\mathbf{k},\sigma'})^2}\,,
\end{aligned}
\label{eq:MT8-causal}
\end{equation}
which corresponds to Eq.~(\textcolor{blue}{7}) of the main text.

Equation~\eqref{eq:MT8-causal} shows that $\chi(\omega)$ receives weight only from inter-branch processes at fixed $\mathbf{k}$ ($\sigma\neq\sigma'$). The identity
\[
\frac{\omega_{\mathbf{k},\sigma}}{\omega_{\mathbf{k},\sigma'}}+\frac{\omega_{\mathbf{k},\sigma'}}{\omega_{\mathbf{k},\sigma}}-2
=\frac{\bigl(\omega_{\mathbf{k},\sigma}-\omega_{\mathbf{k},\sigma'}\bigr)^{2}}{\omega_{\mathbf{k},\sigma}\,\omega_{\mathbf{k},\sigma'}}\,,
\]
makes explicit that the kernel is manifestly non-negative and vanishes quadratically as the two branches approach degeneracy.  Specifically, for $\omega_{\mathbf{k},\sigma'}=\omega_{\mathbf{k},\sigma}+\delta\omega$ with $|\delta\omega|\ll\omega_{\mathbf{k},\sigma}$, the $\mathbf{k}$-resolved weight scales as $\mathcal{O}\!\left((\delta\omega)^2\right)$.
The causal shift $+\mathrm{i}0^+$ in Eq.~\eqref{eq:MT8-causal} enforces analyticity of the retarded susceptibility in the upper half-plane. Using
\begin{equation}
\frac{\Omega}{(\omega+{i}0^+)^2-\Omega^2}
=\frac{1}{2}\!\left[\frac{1}{\omega-\Omega+{i}0^+}-\frac{1}{\omega+\Omega+{i}0^+}\right]\,,
\label{eq:partial}
\end{equation}
where $\Omega=\omega_{\mathbf{k},\sigma}+\omega_{\mathbf{k},\sigma'}$,
together with the Plemelj formula $\mathrm{Im}\,[1/(x+{i}0^+)]=-\pi\delta(x)$, the imaginary part $\ChiPP(\omega)\equiv-\mathrm{Im}\,\ChiR(\omega)$  can be written  as
\begin{equation}
\begin{aligned}
\ChiPP(\omega)=&\frac{\pi\hbar}{8}\sum_{\kk,\sigma\neq\sigma'}
\left|\eps_{\kk,\sigma}^{T} M\,\eps_{-\kk,\sigma'}\right|^2
\left(\frac{\omega_{\kk,\sigma}}{\omega_{\kk,\sigma'}}+\frac{\omega_{\kk,\sigma'}}{\omega_{\kk,\sigma}}-2\right)\\
&\times\left[\delta\!\left(\omega-\omega_{\mathbf{k},\sigma}-\omega_{\mathbf{k},\sigma'}\right)-\delta\!\left(\omega+\omega_{\mathbf{k},\sigma}+\omega_{\mathbf{k},\sigma'}\right)\right].
\label{eq:chiPP-deltas}
\end{aligned}
\end{equation}

Since $\omega_{\mathbf{k},\sigma}+\omega_{\mathbf{k},\sigma'}>0$, for $\omega>0$ only the first term on the right-hand side of Eq.~\eqref{eq:chiPP-deltas} is nonzero,  yielding a nonnegative absorption spectrum, i.e., $\ChiPP(\omega>0)\ge 0$.  
Analyticity of the retarded response in the upper half-plane enforces the
Kramers–Kronig relations, linking the real and imaginary parts of the susceptibility as 
\begin{equation}
\mathrm{Re}\,\ChiR(\omega)=-\frac{1}{\pi}\,\mathcal{P}\!\int_{-\infty}^{\infty}\!d\omega'\,
\frac{\ChiPP(\omega')}{\omega'-\omega}\,,
\label{eq:KK}
\end{equation}
where $\mathcal P$ denotes the Cauchy principal value. 

\paragraph{Finite-Temperature Susceptibility.}

At finite temperature, thermal occupations enter via the normal contraction $\langle \hat a^{\dagger}_{\kk,\sigma} \hat a_{\kk',\sigma'}\rangle = n_{\kk,\sigma}\,\delta_{\kk ,\kk'}\delta_{\sigma,\sigma'}$ with $n_{\kk,\sigma}\equiv (e^{\beta\hbar\omega_{\kk,\sigma}}-1)^{-1}$. These factors modify the Wick expansion and open additional absorption/emission channels in both the anomalous (pairing) and number-conserving sectors.
In a $\mathcal{PT}$-symmetric crystal, the contractions contributing to Eq.~\eqref{eq:Kubo}  are 
\begin{align}
\label{eq:pair_weight}
\Big\langle\big[\hat a_{\kk,\sigma}\hat a_{-\kk\sigma'},\,\hat a_{-\kk,\sigma'}^\dagger \hat a_{\kk,\sigma}^\dagger\big]\Big\rangle_T
&=1+n_{\kk,\sigma}+n_{\kk,\sigma'} \,,\\
\label{eq:nc_weight}
\Big\langle\big[\hat a_{\kk,\sigma}^\dagger \hat a_{\kk,\sigma'},\,\hat a_{\kk,\sigma'}^\dagger \hat a_{\kk,\sigma}\big]\Big\rangle_T
&=n_{\kk,\sigma}-n_{\kk,\sigma'}\,.
\end{align}

Plugging Eqs.~\eqref{eq:pair_weight} and \eqref{eq:nc_weight} into, respectively, Eq.~\eqref{Chia} and \eqref{Chinc},  one finds  

\begin{equation}
\begin{aligned}
&\chi _{\rm a}(\omega,T)
=\frac{\hbar}{4}\sum_{\mathbf k,\sigma\neq\sigma'}
\big|\varepsilon^T_{\mathbf k,\sigma}M\,\varepsilon_{-\mathbf k,\sigma'}\big|^2
\Big(\frac{\omega_{\mathbf k,\sigma}}{\omega_{\mathbf k,\sigma'}}+\frac{\omega_{\mathbf k,\sigma'}}{\omega_{\mathbf k,\sigma}}-2\Big)
\\
&\times
\big(1+n_{\mathbf k,\sigma}+n_{\mathbf k,\sigma'}\big)\frac{\omega_{\mathbf k,\sigma}+\omega_{\mathbf k,\sigma'}}{(\omega+i0^+)^2-(\omega_{\mathbf k,\sigma}+\omega_{\mathbf k,\sigma'})^2} \,,
\label{ChiaT}
\end{aligned}
\end{equation}
and

\begin{equation}
\begin{aligned}
&\chi _{\rm nc}(\omega,T)
=\frac{\hbar}{4}\sum_{\mathbf k,\sigma\neq\sigma'}
\big|\varepsilon^\dagger_{\mathbf k,\sigma}M\,\varepsilon_{\mathbf k,\sigma'}\big|^2
\Big(\frac{\omega_{\mathbf k,\sigma}}{\omega_{\mathbf k,\sigma'}}+\frac{\omega_{\mathbf k,\sigma'}}{\omega_{\mathbf k,\sigma}}+2\Big)
\\[-1mm]
&\times \big(n_{\mathbf k,\sigma}-n_{\mathbf k,\sigma'}\big)
\frac{\omega_{\mathbf k,\sigma}-\omega_{\mathbf k,\sigma'}}{(\omega+i0^+)^2-(\omega_{\mathbf k,\sigma}-\omega_{\mathbf k,\sigma'})^2}\,.
\label{ChincT}
\end{aligned}
\end{equation}
For $\omega>0$, the imaginary parts of Eqs. \eqref{ChiaT} and \eqref{ChincT} are
\begin{equation}
\begin{aligned}
\ChiPP_{\rm a}(\omega,T)&=\frac{\pi\hbar}{8}\sum_{\kk,\sigma\neq\sigma'}
\left|\eps_{\kk,\sigma}^{\TT} M\,\eps_{-\kk,\sigma'}\right|^2
\left(\frac{\omega_{\kk,\sigma}}{\omega_{\kk,\sigma'}}+\frac{\omega_{\kk,\sigma'}}{\omega_{\kk,\sigma}}-2\right)\\
&\times\big(1+n_{\mathbf k,\sigma}+n_{\mathbf k,\sigma'}\big)\delta\!\left(\omega-\omega_{\kk,\sigma}-\omega_{\kk,\sigma'}\right),
\label{eq:chiPP-deltass1}
\end{aligned}
\end{equation}
and
\begin{equation}
\begin{aligned}
\ChiPP_{\rm nc}(\omega,T)&=\frac{\pi\hbar}{4}\sum_{\kk,\sigma\neq\sigma'}
\left|\eps_{\kk,\sigma}^{\dagger} M\,\eps_{\kk,\sigma'}\right|^2
\left(\frac{\omega_{\kk,\sigma}}{\omega_{\kk,\sigma'}}+\frac{\omega_{\kk,\sigma'}}{\omega_{\kk,\sigma}}+2\right)\\
&\times \big(n_{\mathbf k,\sigma}-n_{\mathbf k,\sigma'}\big)
\delta\!\left(\omega-\omega_{\kk,\sigma}+\omega_{\kk,\sigma'}\right).
\label{eq:chiPP-deltass}
\end{aligned}
\end{equation}

\section{S2. Spin relaxation from $J_z^{\mathrm{ph}}$ coupling}

The Hamiltonian of a two-level system interacting with the \(z\)-component of the phonon bath’s total angular momentum can be generally written as
\begin{equation}
\hat H=\tfrac{\hbar\omega}{2}\,\hat\sigma_z+\lambda\,\hat\sigma^{+}\,\hat J_z^{\mathrm{ph}}+\text{h.c.},
\end{equation}
where \(\omega\) is the two-level splitting, \(\lambda\) parameterizes the spin–phonon coupling strength, and \(\hat J_z^{\mathrm{ph}}\) is the bath’s total angular momentum along \(z\). The ladder operators  \(\hat\sigma^{\pm}\equiv(\hat\sigma_x\pm i\hat\sigma_y)/2\) act on the excited state $|e\rangle$ and ground state $|g\rangle$ as   $\hat\sigma^{+}|g\rangle=|e\rangle $and \(\hat\sigma^{-}|e\rangle=|g\rangle\). 
The transition amplitude from the initial state \(|i\rangle=|e\rangle\!\otimes\!|0\rangle\) to the final state \(|f\rangle=|g\rangle\!\otimes\!|n\rangle\), where \(|n\rangle\) is a phonon eigenstate of the uncoupled bath, can be written as
\begin{equation}
\begin{aligned}
c_f(t)&=-\frac{i}{\hbar} \int_0^t d t^{\prime}\langle f| \hat H_{\mathrm{int}}\left(t^{\prime}\right)|i\rangle\\
&=-\frac{i}{\hbar} \lambda^* \int_0^t d t^{\prime} e^{-i \omega t^{\prime}}\langle n| \hat{J}^{\mathrm{ph}}_z\left(t^{\prime}\right)|0\rangle .
\end{aligned}
\label{cf}
\end{equation}

Using Eq.~\eqref{cf},  the transition probability reads is
\begin{equation}
\begin{aligned}
P_{|e\rangle \rightarrow |g\rangle}(t)=\sum_n\left|c_f(t)\right|^2&=\frac{|\lambda|^2}{\hbar^2} \int_0^t d t^{\prime} \int_0^t d t^{\prime \prime} e^{-i \omega\left(t^{\prime}-t^{\prime \prime}\right)}\\
&\times\langle 0| \hat{J}^{\mathrm{ph}}_z\left(t^{\prime \prime}\right) \hat{J}^{\mathrm{ph}}_z\left(t^{\prime}\right)|0\rangle  \,,
\end{aligned}
\end{equation}
where the sum runs over all phonon  eigenstates \(|n\rangle\). Since $\hat{J}_z^{\mathrm{ph}}$ is bilinear in phonon operators,  only two-phonon final states $|n\rangle\sim\hat a^\dagger_{\kk,\sigma}\hat a^\dagger_{-\kk,\sigma'}|0\rangle$ contribute to relaxation. 

In the long-time (i.e., Markov) limit, i.e., $t\rightarrow\infty$,  the relaxation rate can be written as  
\begin{equation}
\begin{aligned}
\Gamma_{|e\rangle \rightarrow |g\rangle}\left(\omega\right)&=\frac{|\lambda|^2}{\hbar^2} \int_{-\infty}^{+\infty} d t e^{i \omega t}\langle0|\hat{J}^{\mathrm{ph}}_z(t) \hat{J}^{\mathrm{ph}}_z(0)|0\rangle \\
&=\frac{|\lambda|^2}{\hbar^2}S_{J_z J_z}(\omega)\,,
\end{aligned}
\label{Gamma}
\end{equation}
where
\begin{equation}
S_{J_z J_z}(\omega) \equiv \int_{-\infty}^{+\infty} d t e^{i \omega t}\langle0|\hat{J}^{\mathrm{ph}}_z(t) \hat{J}^{\mathrm{ph}}_z(0)|0\rangle \,,
\end{equation}
is the unsymmetrized noise spectrum.
Invoking $\chi^{\prime \prime}(\omega)=\frac{1}{2 \hbar}\left[S_{J_z J_z}(\omega)-S_{J_z J_z}(-\omega)\right]$, with  $S_{J_z J_z}(-\omega)=e^{-\beta \hbar \omega} S_{J_z J_z}(\omega)=0$  at zero temperature, Eq.~\eqref{Gamma} can be written as

\begin{equation}
\Gamma(\omega)=\Gamma_{|e\rangle \rightarrow |g\rangle}\left(\omega \right)=\frac{2|\lambda|^2}{\hbar}\ChiPP(\omega)\,.
\label{eq:Gamma}
\end{equation}

For the 2D monoatomic lattice discussed in the main text,  one finds \( |\varepsilon_{\mathbf{k},\sigma}^T M\,\varepsilon_{-\mathbf{k},\sigma'}|^2 = 1-\delta_{\sigma,\sigma'} \), which simplifies Eq.~\eqref{eq:chiPP-deltas} to

\begin{equation}
\begin{aligned}
\ChiPP(\omega)=&\frac{\pi\hbar}{8}\sum_{\kk,\sigma\neq\sigma'}
\left(\frac{\omega_{\kk,\sigma}}{\omega_{\kk,\sigma'}}+\frac{\omega_{\kk,\sigma'}}{\omega_{\kk,\sigma}}-2\right)\\
&\times\delta\!\left(\omega-\omega_{\kk,\sigma}-\omega_{\kk,\sigma'}\right)\,.
\end{aligned}
\label{simplifiedChi}
\end{equation}

Inserting Eq.~\eqref{simplifiedChi} into Eq.~\eqref{Gamma} yields Eq.~(\textcolor{blue}{8}) of the main text : 

\begin{equation}
\begin{aligned}
\Gamma(\omega)=&\frac{\pi|\lambda|^2}{4}\sum_{\kk,\sigma\neq\sigma'}
\left(\frac{\omega_{\kk,\sigma}}{\omega_{\kk,\sigma'}}+\frac{\omega_{\kk,\sigma'}}{\omega_{\kk,\sigma}}-2\right)\\
&\times \delta\!\left(\omega-\omega_{\kk,\sigma}-\omega_{\kk,\sigma'}\right).
\end{aligned}
\label{simp}
\end{equation}




In the long-wavelength limit, a 2D monatomic lattice is well approximated by two acoustic branches, \(\sigma=L,T\), with isotropic dispersions \(\omega_\sigma(k)=c_\sigma k\). We model the Brillouin zone (BZ) by a Debye disk with cutoff fixed by mode counting, i.e., \(\pi k_D^2=A_{\rm BZ}=(2\pi/a)^2\), 
so the single-phonon cutoffs are $\omega_{D,\sigma}=c_\sigma k_D$. For the two-phonon (sum-frequency) kinematics entering Eq.~\eqref{simp}, the linear theory has support
\[
0<\omega\le \omega_D^{(LT)}\equiv (c_L+c_T)k_D,
\]
and Eq.~\eqref{simp} evaluates to
\begin{equation}
\Gamma(\omega)
\approx\frac{N a^2 |\lambda|^2}{4\,c_L c_T}
\left(\frac{c_L-c_T}{c_L+c_T}\right)^{\!2}\,
\omega\;\Theta(\omega)\,\Theta\!\big(\omega_D^{(LT)}-\omega\big).
\label{eq:GammaDebye}
\end{equation}
Keeping the exact square Brillouin zone rather than the Debye disk only
introduces a smooth angular factor that rounds the spectrum for
$\omega\gtrsim (c_L+c_T)\,\pi/a$ and drives $\Gamma(\omega)\to0$ at the
true kinematic limit $\omega_{\max}=(c_L+c_T)\,\sqrt{2}\,\pi/a$; there is
no divergence at the zone edge. In any realistic lattice, deviations from
strict linearity (mode mixing, band flattening near the BZ boundary) further
regularize the high-frequency behavior without altering the low-$\omega$
scaling in Eq.~\eqref{eq:GammaDebye}.

In the finite temperature case, using Eq.~\eqref{ChiaT} and \eqref{ChincT}, we find that the relaxation rate from the excited to the ground state is given by the generic expression  

\begin{equation}
\begin{aligned}
\Gamma_{|e\rangle \rightarrow |g\rangle}(\omega,T)
&=\frac{2|\lambda|^2}{\hbar}\ChiPP(\omega,T)\,[1+n_{BE}(\omega,T)]\\
&=[1+n_{BE}(\omega,T)]
\frac{\pi|\lambda|^2}{4}\sum_{\kk,\sigma\neq\sigma'}
\left|\eps_{\kk,\sigma}^T M\,\eps_{-\kk,\sigma'}\right|^2
 \\
&\times\Bigg\{\left(\frac{\omega_{\kk,\sigma}}{\omega_{\kk,\sigma'}}+\frac{\omega_{\kk,\sigma'}}{\omega_{\kk,\sigma}}+2\right)
\big(n_{\mathbf k,\sigma}-n_{\mathbf k,\sigma'}\big)\,\\
&\times\delta\!\left(\omega-\omega_{\kk,\sigma}+\omega_{\kk,\sigma'}\right)\\
&+2\left(\frac{\omega_{\kk,\sigma}}{\omega_{\kk,\sigma'}}+\frac{\omega_{\kk,\sigma'}}{\omega_{\kk,\sigma}}-2\right)
\big(1+n_{\mathbf k,\sigma}+n_{\mathbf k,\sigma'}\big)\,\\
&\times \delta\!\left(\omega-\omega_{\kk,\sigma}-\omega_{\kk,\sigma'}\right)
\Bigg\}\,,
\end{aligned}
\label{FullGamma}
\end{equation}
 where the Bose-Einstein factor \(1+n_{\mathrm{BE}}(\omega)\) follows from detailed balance, \(S_{J_z J_z}(-\omega)=e^{-\beta\hbar\omega}\,S_{J_z J_z}(\omega)\). At \(T>0\), thermally activated transitions \(|g\rangle\!\to\!|e\rangle\) also occur, with rates suppressed by \(e^{-\beta\hbar\omega}\), i.e., 
\begin{equation}
\Gamma_{|g\rangle \rightarrow |e\rangle}\left(\omega\right)=\frac{|\lambda|^2}{\hbar^2} S_{J_z J_z}\left(-\omega\right)=e^{-\beta\hbar\omega} \Gamma_{|e\rangle \rightarrow |g\rangle}(\omega,T).
\end{equation}


Comparing Eqs.~\eqref{FullGamma} and \eqref{simp} shows that at $T=0$ only the anomalous channel survives: the signal is confined to the sum frequency $\omega_{\mathbf{k},\sigma}+\omega_{\mathbf{k},\sigma'}$, while the number-conserving channel -- responsible for the difference frequency $\left|\omega_{\mathbf{k},\sigma}-\omega_{\mathbf{k},\sigma'}\right|$ -- vanishes identically.

To pin down the frequency support of the number-conserving and anomalous parts of the phonon angular-momentum response, we use the Heisenberg equation $\,\dot{\hat O}=(i/\hbar)[\hat H,\hat O]\,$. The relevant bilinear commutators are
\begin{align}
[\hat H,\hat a^\dagger_{\kk,\sigma}\hat a_{\kk,\sigma'}]&=\hbar(\omega_{\kk,\sigma}-\omega_{\kk,\sigma'})\,\hat a^\dagger_{\kk,\sigma}\hat a_{\kk,\sigma'} \,,\\
[\hat H,\hat a^\dagger_{\kk,\sigma}\hat a^\dagger_{-\kk,\sigma'}]&=\hbar(\omega_{\kk,\sigma}+\omega_{\kk,\sigma'})\,\hat a^\dagger_{\kk,\sigma}\hat a^\dagger_{-\kk,\sigma'} \,,
\end{align}
which imply   
\begin{align}
\hat a^\dagger_{\kk,\sigma}(t)\hat a_{\kk,\sigma'}(t)
&= e^{+i(\omega_{\kk,\sigma}-\omega_{\kk,\sigma'})t}\,
\hat a^\dagger_{\kk,\sigma}\hat a_{\kk,\sigma'} \,,\\
\hat a^\dagger_{\kk,\sigma}(t)\hat a^\dagger_{-\kk,\sigma'}(t)
&= e^{+i(\omega_{\kk,\sigma}+\omega_{\kk,\sigma'})t}\,
\hat a^\dagger_{\kk,\sigma}\hat a^\dagger_{-\kk,\sigma'}.
\end{align}
Invoking the elementary identities
\begin{equation}
\begin{aligned}
(\omega' - \omega)\!\left(\sqrt\frac{\omega'}{\omega} + \sqrt\frac{\omega}{\omega'}\right) &=(\omega' + \omega)\!\left(\sqrt\frac{\omega'}{\omega} - \sqrt\frac{\omega}{\omega'}\right) \\
&= \frac{\omega'^2 - \omega^2}{\sqrt{\omega\,\omega'}} \,,
\end{aligned}
\end{equation}
we can then rewrite
\begin{equation}
\begin{aligned}
[\hat J_{\rm nc},\hat H]=&\frac{\hbar^2}{2}\!\sum_{\mathbf k,\sigma\neq\sigma'}\frac{\omega_{\mathbf k,\sigma'}^2 - \omega_{\mathbf k,\sigma}^2}{\sqrt{\omega_{\mathbf k,\sigma}\,\omega_{\mathbf k,\sigma'}}}
\varepsilon^\dagger_{\mathbf k,\sigma}M\varepsilon_{\mathbf k,\sigma'}\,\\
&\times \hat a^\dagger_{\mathbf k,\sigma}\hat a_{\mathbf k,\sigma'} \,,
\label{Jnc-final}
\end{aligned}
\end{equation}
and 
\begin{equation}
\begin{aligned}
[\hat J_{\rm a},\hat H]
=&\frac{\hbar^2}{2}\!\sum_{\mathbf k,\sigma\neq\sigma'}
\frac{\omega_{\mathbf k,\sigma’}^2 - \omega_{\mathbf k,\sigma}^2}{\sqrt{\omega_{\mathbf k,\sigma}\,\omega_{\mathbf k,\sigma’}}}
\varepsilon^T_{\mathbf k,\sigma}M\varepsilon_{-\mathbf k,\sigma'}\\
&\times\left(\frac12\hat a^\dagger_{\mathbf k,\sigma}\hat a^\dagger_{-\mathbf k,\sigma'}
-\frac12\hat a_{\mathbf k,\sigma}\hat a_{-\mathbf k,\sigma'}\right).
\label{eq:Ja-final}
\end{aligned}
\end{equation}
Equations~\eqref{Jnc-final} and \eqref{eq:Ja-final}, which add up to Eq.~(\textcolor{blue}{9}) of the main text, show clearly that the difference factor $(\omega'-\omega)$ is embedded in the phase of the number-conserving bilinear, whereas the sum factor $(\omega'+\omega)$ resides in the phase of the anomalous pair bilinear; both contributions come multiplied by the same amplitude prefactor $(\omega'^2-\omega^2)/(\omega\,\omega')$ set by the polarization matrix element.
Accordingly, Eq.~\eqref{eq:chiPP-deltass1} 
 -- stemming from fluctuations of the anomalous (pair) sector of the phonon angular momentum  -- has support exclusively on the two–phonon (sum–frequency) shell, whereas, at finite \(T\), the number-conserving channel ~\eqref{eq:chiPP-deltass} gains weight \(\propto n_{\mathbf k,\sigma}-n_{\mathbf k,\sigma'}\)  and is supported at the difference frequency \(\omega=\lvert\omega_{\mathbf k,\sigma}-\omega_{\mathbf k,\sigma'}\rvert\).

\section{S3. Classical analogy}

In this section,  we show that the expectation value \( \langle \hat J^{\rm ph}_z(t) \rangle \) in a suitably prepared two-mode coherent state  reproduces Eq.~(\textcolor{blue}{11}) of the main text. 
Focusing on a 2D monatomic square lattice and excite two modes at fixed \(\mathbf k=(k,0)\) along \(\hat x\): one on branch \(\sigma=x\) with \(\varepsilon_x=(1,0)\) and frequency \(\omega_y=\omega_x+ \delta\omega\), and one on branch \(\sigma'=y\) with \(\varepsilon_y=(0,1)\) and frequency \(\omega_y\), where \(\delta\omega>0\). Using mass-weighted coordinates \(u\equiv \sqrt{m}\,x\), the physical angular momentum \(J^{\rm ph}_z=\sum_{\ell} m\,x_\ell \times \dot{x}_\ell\) becomes \(J^{\rm ph}_z=\sum_{\ell} u_\ell^{T}(iM)\,\dot{u}_\ell\). We prepare a product of coherent states on \(\pm\mathbf k\) for each branch:
\begin{equation}
\begin{aligned}
&|\psi\rangle=\Biggl(\bigotimes_{s=\pm}|{\alpha_x}\rangle_{s\mathbf{k},x}\Biggr)
\otimes
\Biggl(\bigotimes_{s=\pm}|{\alpha_y}\rangle_{s\mathbf{k},y}\Biggr),\\
&\alpha_x=\sqrt{\frac{N \omega_x}{2 \hbar}}\,u_0,\quad \alpha_y=\sqrt{\frac{N \omega_y}{2 \hbar}}\,u_0,
\quad
\alpha_{-\mathbf{k},\sigma}=\alpha_{\mathbf{k},\sigma}^*,
\end{aligned}
\end{equation}
while imposing the condition $\alpha_{-\mathbf k,\sigma}=\alpha^*_{\mathbf k,\sigma}$, such that $\langle \hat{\mathbf u}(t)\rangle$ matches the classical two–mode displacement used in Eq.~(\textcolor{blue}{10}) of the main text. 
The nonvanishing bilinear means are
\begin{equation}
\begin{aligned}
&\big\langle \hat a^\dagger_{\kk,\sigma}(t)\hat a_{\kk,\sigma'}(t)\big\rangle
=\alpha_\sigma^*\alpha_{\sigma'}\,e^{+i(\omega_\sigma-\omega_{\sigma'})t},\\
&\big\langle \hat a^\dagger_{\kk,\sigma}(t)\hat a^\dagger_{-\kk,\sigma'}(t)\big\rangle
=\alpha_\sigma^*\alpha_{\sigma'}^*\,e^{+i(\omega_\sigma+\omega_{\sigma'})t}.
\label{eq:bilinear-means}
\end{aligned}
\end{equation}
Inserting Eq.~\eqref{eq:bilinear-means} into Eq.~(\textcolor{blue}{3}),  one obtains
\begin{align}
\langle \hat J^{\rm nc}_z(t)\rangle&=-\frac{N u_0^2}{2}\,(\omega_x+\omega_y)\,\sin(\delta\omega\,t),
\\
\langle \hat J^{\rm a}_z(t)\rangle&=-\frac{N u_0^2}{2}\,\delta\omega\,\sin\!\big[(\omega_x+\omega_y)t\big]\,,
\end{align}
and therefore
\begin{equation}
\langle \hat J^{\rm ph}_z(t)\rangle
=-\,\frac{N\,u_0^2}{2}\Big[\delta\omega\sin\!\big(\omega_x+\omega_y)t
+(\omega_x+\omega_y)\sin\delta\omega\,t\Big].
\label{S5.6}
\end{equation}

For unit-cell normalization $N=1$ and the amplitude convention adopted in Eq.~(\textcolor{blue}{10}), Eq.~\eqref{S5.6}
coincides exactly with Eq.~(\textcolor{blue}{11}) of the main text: a fast sum-frequency carrier modulated by a slow difference-frequency envelope with zero time average over  $2\pi/\delta\omega$.
The same result follows either by (i) taking the classical limit of the commutator Eq.~(\textcolor{blue}{9}) in the main text, i.e., $\frac{d}{dt} O_{\rm cl}(t)=\{O_{\rm cl}(t),H_{\rm cl}\}$, or (ii) evaluating the commutator Eq.~(\textcolor{blue}{9}) on the coherent state.  

\section{S4. Angular momentum fluctuation and absorption rate in a 3D lattice}

In this section, we illustrate how the general susceptibility framework developed above manifests in a realistic three-dimensional lattice where two-phonon excitations can be generated experimentally. A particularly clean route is resonant two-phonon Raman scattering, in which a circularly polarized pump near an electronic resonance coherently creates pairs of phonons with opposite momenta \((\kk,-\kk)\). When the two branches involved are nondegenerate and carry a finite angular momentum form factor $F_{\kk,\sigma,\sigma'}$, their superposition produces instantaneous lattice angular momentum oscillations at the beating frequency \(\delta \omega=\lvert \omega_{\kk,\sigma}-\omega_{\kk,\sigma'}\rvert\). To make this concrete, we consider diamond silicon (space group $F d \overline{3} m$), which preserves both inversion $\mathcal{P}$ and time-reversal $\mathcal{T}$.
Under indirect-gap resonance, the two-phonon Raman amplitude selectively enhances phonon pairs with wavevectors near X. We thus focus on the \(\pm X\) LO and TO phonons, with \(\hbar\omega_{\mathrm{TO}}(\pm X)\approx 58~\mathrm{meV}\) and \(\hbar\omega_{\mathrm{LO}}(\pm X)\approx 64~\mathrm{meV}\). If their angular momentum form factor is nonzero, their coherent superposition produces instantaneous lattice angular momentum beating at \(\delta\omega\approx 1.6~\mathrm{THz}\). We proceed to assess whether this requirement is met.

We consider the diamond lattice with two atoms per unit cell, \(A,B\), at relative position \(\tfrac a4(1,1,1)\), where $a$ is the lattice constant. At the Brillouin–zone edge \(X=(2\pi/a,0,0)\), symmetry enforces three twofold degeneracies: 

\begin{align}
&\omega_{X,LO}=\omega_{X,LA}\equiv\omega_{LO},\\
&\omega_{X,TO_1}=\omega_{X,TO_2}\equiv\omega_{TO},\\
&\omega_{X,TA_1}=\omega_{X,TA_2}\equiv\omega_{TA}.
\label{omegalo}
\end{align}

Fixing a smooth gauge for the polarization vectors that varies continuously as $\mathbf{k}$ is displaced from $X$, one convenient choice is
\begin{equation}
\begin{array}{ll}
\mathrm{LA}: \varepsilon_{\mathrm{LA}}=\frac{1}{\sqrt{2}}\left(\hat{x}_A,\hat{x}_B\right), & \mathrm{LO}: \varepsilon_{\mathrm{LO}}=\frac{1}{\sqrt{2}}\left(\hat{x}_A,-\hat{x}_B\right), \\
\mathrm{TA}_1: \varepsilon_{\mathrm{TA} 1}=\frac{1}{\sqrt{2}}\left(\hat{y}_A,\hat{y}_B\right), & \mathrm{TA}_2: \varepsilon_{\mathrm{TA} 2}=\frac{1}{\sqrt{2}}\left(\hat{z}_A,\hat{z}_B\right), \\
\mathrm{TO}_1: \varepsilon_{\mathrm{TO} 1}=\frac{1}{\sqrt{2}}\left(\hat{y}_A,-\hat{y}_B\right), & \mathrm{TO}_2: \varepsilon_{\mathrm{TO} 2}=\frac{1}{\sqrt{2}}\left(\hat{z}_A,-\hat{z}_B\right) ,
\end{array}
\label{varepsilon}
\end{equation}
where  $\hat x,\hat y,\hat z$ are unit vectors along Cartesian axes and $A,B$ denote the two atoms in the primitive cell.
With the $z$-component of angular momentum represented by
\begin{equation}
M=\begin{pmatrix}0&-i&0\\[2pt]i&0&0\\[2pt]0&0&0\end{pmatrix},
\end{equation}
the only nonvanishing $X$-point form factors entering Eq.~\eqref{eq:kernel-def} are
\begin{equation}
\begin{aligned}
&|F_{X,\rm LO,TO_1}|^2 =\frac 14 \left(\frac{\omega_{LO}}{\omega_{TO}}+\frac{\omega_{TO}}{\omega_{LO}}-2\right),\\
&|F_{X,\rm LA,TA_1}|^2 =\frac 14 \left(\frac{\omega_{LA}}{\omega_{TA}}+\frac{\omega_{TA}}{\omega_{LA}}-2\right),\\
\end{aligned}
\label{FinSi}
\end{equation}
making positivity explicit and showing that the couplings vanish only at degeneracy ($\omega_{\mathrm{LO}}=\omega_{\mathrm{TO}}$ or $\omega_{\mathrm{LA}}=\omega_{\mathrm{TA}}$). Because the gauge in Eq.~\eqref{varepsilon} is smooth, Eq.~\eqref{FinSi} remains a good approximation in a finite neighborhood of $X$. The nonzero form factors~\eqref{FinSi} demonstrate that LO and TO modes at $\pm X$ sustain angular momentum coherence; their superposition yields a beat at $\delta\omega = \lvert \omega_{X,LO} - \omega_{X,TO} \rvert$. In the time-resolved resonant Raman geometry, the resulting dynamical lattice angular momentum can drive the antisymmetric Raman tensor and manifest as a transient polarization rotation/ellipticity of the probe.

\bibliography{references}